\documentclass[aps,prd,nofootinbib,superscriptaddress,twocolumn]{revtex4-1}

\usepackage{amsmath}
\usepackage{amssymb}
\usepackage{bbold}
\usepackage{epsfig}
\usepackage{graphicx}
\usepackage{float} 
\usepackage{color}
\usepackage{xcolor}
\usepackage{nicefrac}
\usepackage{epstopdf}
\usepackage{tabularx}
\usepackage[colorlinks=true, pdfstartview=FitV, linkcolor=blue, citecolor=purple,urlcolor=blue]{hyperref}
\definecolor{darkgreen}{rgb}{0,0.5,0}
\definecolor{purple}{rgb}{0.5,0,0.5}
\definecolor{nblue}{rgb}{0.0,0.0,0.50}
\definecolor{scarlet}{rgb}{1.0,0.2,0}

\begin{document}

\title{Charm and strange meson fragmentation functions }

\author{Roberto C. da Silveira}
\email{rc.silveira@unesp.br}
\affiliation{Instituto de F\'{\i}sica Te\'orica, Universidade Estadual Paulista, Rua Dr.~Bento Teobaldo Ferraz 271 - Bloco II, 01140-070 S\~ao Paulo, S\~ao Paulo, Brazil}

\author{Ian C. Clo\"{e}t}
\email{icloet@anl.gov}
\affiliation{Physics Division, Argonne National Laboratory, Lemont, Illinois 60439, USA}

\author{Bruno El-Bennich}
\email{bennich@unifesp.br}
\affiliation{Instituto de F\'{\i}sica Te\'orica, Universidade Estadual Paulista, Rua Dr.~Bento Teobaldo Ferraz 271 - Bloco II, 01140-070 S\~ao Paulo, S\~ao Paulo, Brazil}
\affiliation{Departamento de F\'isica, Universidade Federal de S\~ao Paulo, Rua S\~ao Nicolau 210, 09913-030 Diadema, S\~ao Paulo, Brazil}

\author{Fernando E. Serna}
\email{feserna@uninorte.edu.co}
\affiliation{Departamento de Matem\'aticas y Estad\'istica, Universidad del Norte, Km 5 V\'ia Antigua Puerto Colombia, Barranquilla 081007, Colombia}


\begin{abstract}
Quark fragmentation functions describe the hadronization process of a quark where any of the final-state hadrons carries a fraction of its initial momentum. We compute these fragmentation functions for a cascade that includes pions, kaons, and the charmed $D$ and $D_s$ mesons, starting from the elementary quark-to-meson fragmentation process. The latter is obtained from the relevant cut diagram, and employs Poincar\'e covariant Bethe-Salpeter wave functions and quark propagators. We derive a set of twenty-five coupled jet equations that describe the cascade of emitted mesons in the fragmentation process. Their solutions yield full fragmentation functions that offer a consistent picture of the quark fragmentations across the light and heavy sectors.
\end{abstract}

\date{\today}
\maketitle


\section{Introduction}
\label{intro}

Fragmentation functions are fundamental objects in the description of hadron production with mostly longitudinal momenta and relatively small transverse momenta
in high-energy processes. They encode the nonperturbative dynamics by which quarks and gluons transform into jets of color-neutral hadrons, and appear as key inputs in 
factorization theorems for processes such as electron-positron annihilation, semi-inclusive deep inelastic scattering, and proton--proton collisions~\cite{Anselmino:2008jk,
Ellis:2008in,Hirai:2007ww}. Among them, the function $D_q^h(z)$  represents the probability density for a quark  $q$ to produce a hadron $h$ carrying a fraction $z$  
of its light-front momentum. Despite being essential for phenomenology, fragmentation functions remain difficult to compute from first principles, due to their inherently nonperturbative and timelike nature~\cite{Field:1976ve,Collins:1981uw,Metz:2016swz} and much effort is dedicated to global analysis to extract them and their uncertainties from data~\cite{deFlorian:2007aj,
deFlorian:2014xna,deFlorian:2017lwf,Borsa:2022vvp,AbdulKhalek:2022laj,Ochoa-Oregon:2023ktx}.

Low-energy effective models have been used to study fragmentation functions, in particular within the context of light-meson jet functions~\cite{Ito:2009zc,Matevosyan:2011vj,Casey:2012ux,
Matevosyan:2010hh,Xing:2023pms,Xing:2025eip,daSilveira:2024ddq,Bopsin:2025vhz}. However, most existing approaches either rely on point-like approximations for the 
hadron structure or neglect the internal dynamics of the bound states. Moreover, charmed fragmentation functions which involve the  $D$ and $D_s$ mesons  are often 
introduced via phenomenological fits or through factorized perturbative expressions involving hadron distribution amplitudes~\cite{Cacciari:1996wr,Kneesch:2007ey}. 
A consistent, fully covariant description of fragmentation that treats both light and heavy mesons on the same footing has remained challenging.

We compute the elementary fragmentation functions of pseudoscalar mesons, extending them from pions and kaons, to the $D$ and $D_s$ mesons, within a Poincar\'e covariant 
framework  based on the Bethe-Salpeter equation (BSE). Starting from an elementary quark-to-meson fragmentation process, we construct a kernel for a set of coupled 
integral equations that models the cascade of hadron emissions in a jet. The elementary fragmentation function is related to the quark distribution function via crossing and charge
symmetry~\cite{Drell:1969jm, Drell:1969wd, Drell:1969wb} and makes use of dressed quark propagators and the  Bethe-Salpeter amplitude (BSA) of the outgoing 
mesons. We then solve the jet functions that describe the cascade of fragmentation into hadrons, incorporating contributions from all intermediate channels and 
enforcing the normalization constraints implied by momentum conservation.

This method allows us to compute the full fragmentation functions $D_q^h (z)$ for a variety of quark flavors and mesons, without resorting to external parameterizations or 
models of the vertex functions. We remind that  $D_q^h(z,\mu) dz$ can be interpreted as a probability distribution that at a given energy scale $\mu$ a quark of flavor $q$ 
escaping the collision region produces a hadron $h$, thereby losing a light-front fraction, $z$, of its original momentum. The resulting FF exhibit the expected suppression 
of heavy meson production from light quarks and show characteristic features of charm fragmentation, such as the dominance of the $c \to D$ and $c \to D_s$ channels 
at intermediate-to-high value of $z$. Our approach provides a unified and consistent description of quark hadronization, and offers a basis for future applications to include 
vector mesons and nucleons in the future.


\section{Dressed quark propagators}

The description of quark dynamics at low momenta necessitates nonperturbative methods. Central to this framework is the fully dressed quark propagator, which encodes 
essential information about confinement and dynamical chiral symmetry breaking (DCSB). This quark satisfies a quark gap equation, which in a functional
approach is given by a Dyson-Schwinger equation (DSE) that self-consistently incorporates quark and gluon dressing effects and represents an infinite tower 
of integral equations~\cite{Bashir:2012fs,Cloet:2013jya} for a given flavor $f$, 
\begin{align}
S^{-1}_f (p) &= Z_2^f \left(i\gamma\cdot p + m_{\rm bm}^f \right) \nonumber \\
+ Z_1^f & g^2 \! \int^\Lambda \!\! \frac{d^4k}{(2\pi)^4} \, D^{ab}_{\mu\nu}(q) \, \frac{\lambda^a}{2} \gamma_\mu \, S_f(k)\, \Gamma^b_{\nu}(k,p) \, ,
\label{eq:DSEquark}
\end{align}
where $q = k - p$, $m_{\rm bm}^f$ is the bare current-quark mass, and $Z_{1,2}^f (\mu\Lambda)$ are the renormalization constants for the quark-gluon vertex and wave 
function determined at the scale~$\mu$. The gluon propagator $D^{ab}_{\mu\nu}(q)$ and quark-gluon vertex $\Gamma^b_{\nu}(k,p)$~\cite{Albino:2018ncl,
Albino:2021rvj,El-Bennich:2022obe,Lessa:2022wqc} are both dressed, incorporating strong interaction effects that are inaccessible in perturbation theory.
The regularization scale $\Lambda$ of the Euclidean integral and can be taken to infinity, and in practice $\Lambda \gg \mu$.

The general solution of Eq.~\eqref{eq:DSEquark} admits a Poincar\'e covariant decomposition in terms of vector and scalar dressing functions,
\begin{align}
\hspace*{-2mm}
S_f (p) = -i\gamma\cdot p \, \sigma_v^f (p^2) + \sigma_s^f(p^2) = \frac{Z_f(p^2)}{i\gamma\cdot p + M_f(p^2)} \, ,
\label{eq:DressedProp}
\end{align}
in which $Z(p^2)$ is the momentum-dependent wave function renormalization and $M(p^2)$ the running mass function. These functions are determined self-consistently 
imposing the renormalization conditions,
\begin{align}
Z_f(\mu^2) &= 1 \, , \\
S^{-1}_f(\mu^2) &= i\gamma\cdot p + m_f(\mu) \, ,
\end{align}
where $m_f(\mu)$ is the renormalized mass at scale $\mu$, related to the bare mass via the mass renormalization constant $Z_4^f(\mu,\Lambda)$:
\begin{equation}
Z_4^f(\mu,\Lambda) \, m_f (\mu) = Z_2^f(\mu,\Lambda)\, m_{\rm bm}^f (\Lambda) \, .
\end{equation}

A widely used, successful and symmetry-preserving approximation to the full DSE system is the rainbow-ladder truncation, which is implemented  by restricting
the fully dressed quark gluon vertex to the leading Dirac component: $\Gamma_{\nu, f} \rightarrow Z_2^f \gamma_\nu$. Allowing for a flavor dependence of the
interaction, the DSE kernel becomes~\cite{Serna:2018dwk},
\begin{equation}
Z_1^f g^2 \!D_{\mu\nu}(q)\, \Gamma^b_{\nu}(k,p) \rightarrow  \big (Z_2^f \big)^{\!2} \mathcal{G}_f(q^2) D_{\mu\nu}^{\rm free}(q) \frac{\lambda^a}{2} \gamma_\nu \, ,
\label{eq:RLkernel}
\end{equation}
where $\mathcal{G}(q^2)$ is an effective interaction modeling the combined dressing effects of the gluon propagator and the quark-gluon vertex. 
The Landau-gauge free gluon propagator,
\begin{equation}
D_{\mu\nu}^{\rm free}(q) = \delta^{ab} \left( \delta_{\mu\nu} - \frac{q_\mu q_\nu}{q^2} \right) \frac{1}{q^2} \, ,
\end{equation}
ensures transversality.

The flavor-dependent dressing function $\mathcal{G}(q^2)$ is a variation of an infrared massive and finite interaction consistent with predictions of 
contemporary  DSE and lattice-QCD studies of the gluon dressing function~\cite{Qin:2011dd}. It consists of an  infrared-dominant term and a 
perturbative tail that ensures the correct ultraviolet behavior:
\begin{equation}
\frac{\mathcal{G}_f(q^2)}{q^2} = \mathcal{G}^\mathrm{IR}_f (q^2) + 4\pi\, \tilde{\alpha}_{\rm PT}(q^2) \, .
\label{eq:Gfmodel}
\end{equation}
Following Ref.~\cite{Qin:2011dd}, the infrared part is parameterized as:
\begin{equation}
\mathcal{G}_f^\mathrm{IR}(q^2) = \frac{8\pi^2 D_f}{\omega_f^4} \, \exp\left( -\frac{q^2}{\omega_f^2} \right) \, ,
\end{equation}
while the ultraviolet term reads:
\begin{equation}
4\pi\, \tilde{\alpha}_{\rm PT}(q^2) = \frac{8\pi^2\, \gamma_m\, \mathcal{F}(q^2)}{\ln\left[ \tau + \left(1 + q^2/\Lambda^2_{\rm QCD} \right)^2 \right]} \, ,
\end{equation}
with $\mathcal{F}(q^2) = [1 - \exp(-q^2/4m_t^2)]/q^2$, $\gamma_m = 12/(33 - 2N)$, $m_t = 0.5$~GeV, $\tau = e^2 - 1$, and $\Lambda_{\rm QCD} = 0.234$~GeV. Values of the interaction parameters, $\omega_f$ and $D_f$, the renormalized masses, $m_f$, and of the renormalization constants, $Z_2^f$ and $Z_4^f$, at $\mu=2$\;GeV are found in Tab.~1 of Ref.~\cite{Serna:2020txe}.


\section{Pseudoscalar Meson Bound States}

The structure and properties of pseudoscalar mesons are described by solutions of the homogeneous BSE, which provides a Poincar\'e covariant description of 
quark-antiquark bound states. For a meson, the BSE reads~\cite{Llewellyn-Smith:1969bcu}:
\begin{equation}
  \Gamma_m (k, p) =  \int^\Lambda \!  \frac{d^4q}{(2\pi)^4} \,  K(k,q,p) \, \chi_m  (q,p)  \,,
\label{BSE-ps}
\end{equation}
where $k$ is the relative momentum between the quark and antiquark, $p$ is the total momentum of the bound state, and $K(k,q,p)$ is the fully amputated scattering kernel. 
The Bethe-Salpeter wave function is defined as the BSA with attached dressed quark propagators,
\begin{equation}
   \chi_m  (k,p) = S(k_\eta)\, \Gamma_m (k,p) S(k_{\bar\eta}) \,,
\end{equation}
in which $k_\eta = k + \eta p$ and  $k_{\bar\eta} = k - \bar\eta p$ involve momentum-partitioning parameters $\eta$ and $\bar \eta$, such that $\eta + \bar\eta = 1$.

The general Poincar\'e-covariant form of the pseudoscalar meson's BSA consistent with $J^{PC} = 0^{-+}$ can be expressed as,
\begin{align}
   \Gamma_m (k,  &\, p) = \gamma_5 \big[      i E_m (k,p) +    \gamma \cdot p\, F_m (k,p)    \nonumber \\
     + & \  \gamma  \cdot k\, k\cdot p\, G_m (k,p)     + \sigma_{\mu\nu} k_\mu p_\nu \, H_m (k,p) 
   \big] \,,
 \label{PS-BSA-general}
\end{align}
where $E_m, F_m, G_m, H_m$ are Lorentz-invariant scalar functions. For flavor-symmetric systems such as the neutral pion, charge-symmetry  imposes further 
constraints with respect to $k \cdot p \rightarrow -k \cdot p$ on these amplitudes. In particular, the odd contributions in $k \cdot p$ vanish in this case, leading to a simpler structure of the BSA.

In the calculation of the elementary fragmentation function $d_{q}^{h}(z)$~\eqref{EQ:fragLF}, we employ the BSE solutions of the 
pseudoscalar mesons $\pi$, $K$,$D$, $D_s$ and $\eta_c$. The latter were obtained with the improved ladder kernel~\cite{Serna:2020txe}, 
\begin{equation}
  \label{RLkernel}
    K_{fg}(k,q,p)   = -  \mathcal{Z}_2^2  \  \frac{\mathcal{G}_{fg}  (l^2)}{l^2 } \frac{\lambda^a }{2} \gamma_\nu \frac{\lambda^a }{2} \gamma_\nu \ ,
\end{equation}
where one combines the quark's wave-function renormalization constants, $ \mathcal{Z}_2 (\mu,\Lambda) =\surd Z_2^f\surd Z_2^g$,  and uses
an averaged flavor-dependent interaction, 
\begin{equation}
    \frac{ \mathcal{G}_{fg}  (l^2) }{ l^2 }= \mathcal{G}_{fg}^\mathrm{ IR}(l^2) +  4\pi \tilde\alpha_\mathrm{PT}(l^2)\, .
\end{equation}
The low-momentum domain of the dressing function is modeled as, 
\begin{equation}
\label{BSEflavor}
 \mathcal{G}_{fg}^\mathrm{IR} (l^2)  =   \frac{8\pi^2}{(\omega_f\omega_g)^2} \sqrt{D_f\,D_g }\,e^{-l^2/(\omega_f\omega_g)} \, ,
\end{equation}
and the gluon momentum is $l=k-q$.


\section{Flavored Elementary Fragmentation Functions }

We extend a previous study~\cite{daSilveira:2024ddq} wherein an elementary fragmentation function and related jet functions were limited to the isospin states of
the pion. In doing so, we generalize the formalism to include additional pseudoscalar mesons, namely the kaons and charmed mesons $D$ and $D_s$. 

\begin{figure}[t!] 
\vspace*{-7mm}
\hspace*{-7mm}
\centering
  \includegraphics[scale=0.5,angle=0]{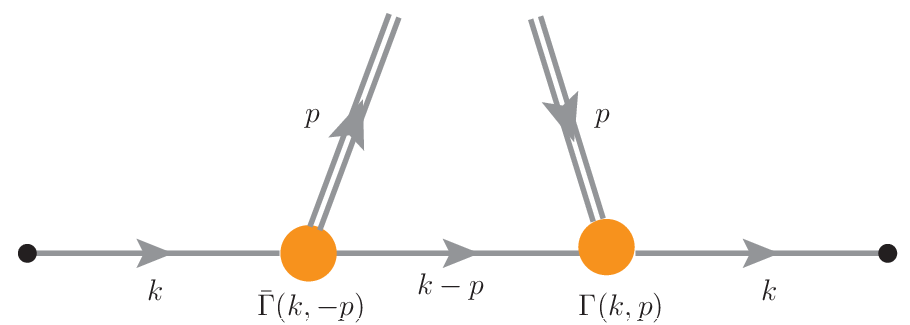} 
\caption{Cut diagram of the fragmentation function $d^{m}_{q}(z)$ in
Eq.~\eqref{EQ:fragLF}. Orange circles with outgoing/incoming double-solid
lines denote the meson in the fragmentation process, solid lines
are quark propagators and the solid dots represent the $\gamma_{+}$
between the two quarks with momentum $k$.}
\label{fig:cut_eff} 
\end{figure}

We begin by computing the elementary fragmentation functions describing the transition of a quark $q$ into a pseudoscalar meson $m$, thereby transferring a 
fraction, $z=k^+/p^+$, of its light-front momentum. The relevant kinematics is  illustrated in Fig.~\ref{fig:cut_eff}. For completeness, we briefly summarize the results 
derived in Ref.~\cite{daSilveira:2024ddq}, omitting technical details found in Ref.~\cite{Ito:2009zc}. We remind that the elementary quark fragmentation function, 
$q\to qm$, for physical $z = 1/x < 1$ is related to the parton distribution function of a meson  for unphysical values $x > 1$ by the  Drell-Levy-Yan (DLY) 
relation~\cite{Drell:1969jm,Drell:1969wd,Drell:1969wb},
\begin{equation}
  d_q^m (z)=\frac{z}{6} f_q^m(x)\, , 
  \label{PDLYrel}
\end{equation}
where the factor $1/6$ is due to the spin-color degeneracy of the quark. To a certain extent, Eq.~\eqref{PDLYrel} is useful to relate the kernels of the 
evolution equations of distribution and fragmentation functions at next-to-leading order~\cite{Stratmann:1996hn,Blumlein:2000wh}, though there is no 
reason to assume that the quark fragments into a single meson, as will be further discussed in Sec.~\ref{sec-jet}.

Evaluation of the cut diagram implied by the DLY relation in Fig.~\ref{fig:cut_eff} yields the elementary fragmentation function:
\begin{align} 
   d_{q}^{m} (z)  = & \, \frac{N_c  C_q^m  z}{6}\! \int \! \frac{d^4 k}{(2\pi)^4 } \,   2\pi i\, \delta\Big (k^{+} - \tfrac{p^+}{z} \Big )  \delta\! \left ( \ell^2 + M^2 (\ell^2 ) \right )  
       \nonumber \\
    &   \times   \operatorname{Tr}_\text{D}  \Big [  S(k)\gamma^{+}S(k)\,\bar{\Gamma}_{m} \big (-k+\tfrac{p}{2}, - p \big )  
       \nonumber \\
    &  \times  \big ( -i \gamma \cdot \ell  + M (\ell^2 ) \big ) \, \Gamma_{m} \big (k-\tfrac{p}{2}, p \big )  \Big  ] . 
\label{EQ:fragLF}   
\end{align}
In Eq.~\eqref{EQ:fragLF}, $N_c=3$, $k$ and $\ell = k-p$ are the incoming and outgoing four-momenta of the quark, $S(k)$ denotes dressed quark 
propagators~\eqref{eq:DSEquark},  $\Gamma_{m} (k-\tfrac{p}{2},p)$ and $\bar \Gamma_{m} (-k+\tfrac{p}{2},-p)$ are the BSAs~\eqref{PS-BSA-general}  
and  the charge-conjugate  BSA of the meson with total momentum $p$, respectively. Moreover, $C_q^m$ are isospin factors listed in table~\ref{tab:isospin}. 
The light-front momenta are defined as $k^{+}=n\,\cdot \,k$,  $p^{+}=n\cdot p$ and $\gamma^{+}=\gamma\cdot n$, where $n$ is a light-like four-vector 
defined by $n^2=0$. 

We remind that  Eq.~\eqref{EQ:fragLF} refers to a frame in which the produced meson has no transverse momentum,  $\boldsymbol{p}_T = 0$, while 
the fragmenting quark does possess a momentum $\boldsymbol{k}_T\neq 0$. The interpretation of the fragmentation function as a distribution of 
\emph{mesons in a quark\/} implies that the quark must be boosted to a frame where $\boldsymbol{k}_\perp = 0$, which is achieved with a Lorentz
transformation. The latter yields~\cite{Collins:1981uw,Barone:2001sp},
\begin{equation}
    \boldsymbol{k}_T = -\frac{\boldsymbol{p}_{\perp}}{z} \, . 
\label{boost}    
\end{equation}
where one distinguishes between both frames with the subscripts $T$ and $\perp$, that is, $\boldsymbol{k}_T \neq 0$ but $\boldsymbol{k}_\perp =0$.

\begin{table*}[t!]
\centering 
\renewcommand{\arraystretch}{1.5}
\setlength{\tabcolsep}{6pt}

\begin{tabular}{c|c|c|c|c|c|c|c|c|c|c|c|c|c}
\hline\hline
    $C_q^m$  & $\pi^0$ & $\pi^+$ & $\pi^-$ & $K^0$ & $\overline{K}^0$ & $K^+$ & $K^-$ & $D^0$ & $\overline{D}^0$ & $D^+$ & $D^-$ & $D_s^+$ & $D_s^-$  \\ \hline
    $u$            & 1/2 & 1 & 0 & 0 & 0 & 1 & 0 & 0 & 1 & 0 & 0 & 0 & 0\\
   $d$            & 1/2 & 0 & 1 & 1 & 0 & 0 & 0 & 0 & 0 & 0 & 1 & 0 & 0 \\ 
   $s$            & 0 & 0 & 0 & 0 & 1 & 0 & 1 & 0 & 0 & 0 & 0 & 0 & 1 \\ 
  $c$            & 0 & 0 & 0 & 0 & 0 & 0 & 0 & 1 & 0 & 1 & 0 & 1 & 0 \\
$\overline{u}$ & 1/2 & 0 & 1 & 0 & 0 & 0 & 1 & 1 & 0 & 0 & 0 & 0 & 0 \\
$\overline{d}$ & 1/2 & 1 & 0 & 0 & 1 & 0 & 0 & 0 & 0 & 1 & 0 & 0 & 0 \\
$\overline{s}$ & 0 & 0 & 0 & 1 & 0 & 1 & 0 & 0 & 0 & 0 & 0 & 1 & 0 \\
$\overline{c}$ & 0 & 0 & 0 & 0 & 0 & 0 & 0 & 0 & 1 & 0 & 1 & 0 & 1 \\
\hline\hline
\end{tabular}
\caption{\label{tab:isospin} The isospin/flavor coefficients $C_q^m$ introduced in Eq.~\eqref{EQ:fragLF}.}   
\end{table*} 

Since the parton must fragment into a hadron with unit probability, the normalized elementary fragmentation function $\hat{d}_{q}^{m}$ satisfies, 
\begin{align} 
 \sum_m  \int_{0}^{1}dz\, \hat{d}_{q}^{m} = 1 \, .
\label{EQ:fragNEFF}   
\end{align}
The fragmentation dynamics of light  and heavy  quarks into pseudoscalar mesons involves distinct production channels governed by flavor constraints. 
Specifically, one distinguishes between the cases:
\begin{itemize}
    \item An \emph{up} quark directly fragments into $\pi^+(u\bar{d})$, $\pi^0((u\bar{u}-d\bar{d})/\sqrt{2})$, $K^+(u\bar{s})$ and $\bar{D}^0(u\bar{c})$.
    \item A \emph{down} quark directly fragments into $\pi^-(d\bar{u})$, $\pi^0((u\bar{u}-d\bar{d})/\sqrt{2})$, $K^0(d\bar{s})$ and $D^-(d\bar{c})$.
    \item A \emph{strange} quark produces $K^-(s\bar{u})$, $\bar{K}^0(s\bar{d})$ and $D_s^-(s\bar{c})$.
    \item A \emph{charm} quark fragments into $D^0(c\bar{u})$, $D^+(c\bar{d})$, and $D_s^+(c\bar{s})$.
\end{itemize}
Analogous considerations hold for the antiquarks. 
The normalization conditions for the elementary fragmentation functions must account for all allowed hadronization channels. This leads to the generalized sum rules of Eq.~\eqref{EQ:fragNEFF}:
\begin{align} 
\int_{0}^{1}dz \left[ \tfrac{3}{2}\hat{d}_{u}^{\pi^{+}}(z) + \hat{d}_{u}^{K^{+}}(z) + \hat{d}_{u}^{\bar{D}^{0}}(z)\right] &= 1,  \\
\int_{0}^{1}dz\left[2\hat{d}_{s}^{K^{-}}(z) + \hat{d}_{s}^{D_{s}^{-}}(z)\right] &= 1, \\
\int_{0}^{1}dz\left[2\hat{d}_{c}^{D^{0}}(z) + \hat{d}_{c}^{D_{s}^{+}}(z)\right] &= 1,
\label{EQ:NEFF}   
\end{align}
where the pre-factors account for isospin multiplicities:
\begin{itemize}
    \item The factor $\frac{3}{2}$ combines the $\pi^+$ and $\pi^0$ isospin states of the or $u\to\pi$ fragmentations.  
    \item The factor of 2 in the strange and charm sectors reflects the $K^-/\bar{K}^0$ and $D^0/D^+$ degeneracy.
\end{itemize}


\section{Flavored coupled jet equations\label{sec-jet}}

\begin{figure}[t!]
\centering
  \includegraphics[scale=0.6,angle=0]{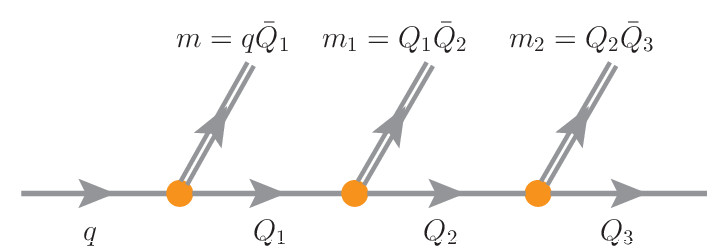} 
\caption{Quark fragmentation cascade process.}
\label{fig:cascade} 
\end{figure}

The fragmentation of quarks into heavy-light mesons beyond the pion is described by recursive integral equations that account for all possible fragmentation pathways 
depicted schematically in Fig.~\ref{fig:cascade}. Assuming isospin symmetry, $m_u = m_d$, and considering pseudoscalar meson valence structures, the full 
fragmentation functions $D_q^m(z)$ are determined through the recursion relation that resums the cascade of elementary fragmentations:
\begin{align}
\hspace*{-1mm}
    D_q^{m}(z) = \hat{d}_q^{m}(z) +\!\!\!  \sum_{Q\in\{u,d,s,c\}} \int_z^1 \frac{dy}{y}  \hat{d}_q^{Q}\left(\frac{z}{y}\right) D_Q^m(y)\, .
\label{EQ:REDF}
\end{align}
The sum runs over all active quark flavors and the integral describes the sequential fragmentation process: an initial quark $q$ produces an intermediate quark 
$Q$ carrying momentum fraction $y$, which then fragments into the meson $m$ with rescaled momentum $z/y$. Therefore, the full fragmentation or jet functions
give the probability for finding a meson $m$ with light-front momentum fraction $z$ in a jet.

The splitting functions $\hat{d}_q^Q(z)$ in Eq.~\eqref{EQ:REDF}, representing the direct emission of a quark $Q$ from quark $q$, are kinematically related 
to the elementary fragmentation functions through via,
\begin{equation}
  \hat{d}_q^Q(z) = \hat{d}_q^m(1-z)\, ,
\end{equation}
%
%
where we consider the pseudoscalar mesons, $\pi^\pm, \pi^0, K^\pm, K^0, \bar{K}^0, D^0, \bar{D}^0, D^\pm$ and $D_s^\pm$ and $Q$ is the 
intermediate quark.  The elementary functions of the $m=q\bar Q$ states are related by $G$-parity~\cite{Lee:1956sw}:
\begin{subequations}
\begin{align}
&  \hspace*{-3mm} \hat{d}_{u}^{u}(z) = \hat{d}_{u}^{\pi^0}(1-z) = \hat{d}_{u}^{\bar \pi^0}(1-z) = \hat{d}_{d}^{d}(z) = \hat{d}_{\bar{u}}^{\bar{u}}(z) ,
\label{EQ:GparityI} \\
&  \hspace*{-3mm}\hat{d}_{u}^{d}(z) = \hat{d}_{u}^{\pi^+}(1-z)= \hat{d}_{d}^{\pi^-}(1-z)  = \hat{d}_{d}^{u}(z) = \hat{d}_{\bar{u}}^{\bar{d}}(z) , \\
&  \hspace*{-3mm}\hat{d}_{u}^{s}(z) = \hat{d}_{u}^{K^+}\!(1-z) = \hat{d}_{u}^{K^0}(1-z) =   \hat{d}_{d}^{s}(z) = \hat{d}_{\bar{u}}^{\bar{s}}(z) , \\
&  \hspace*{-3mm}\hat{d}_{s}^{u}(z) = \hat{d}_{s}^{K^-}\!(1-z) = \hat{d}_{s}^{\bar{K}^0}(1-z) = d_s^d(z)   = d_{\bar s}^{\bar u}(z)  ,   \\
&  \hspace*{-3mm}\hat{d}_{c}^{u}(z) = \hat{d}_{c}^{D^0}(1-z) = \hat{d}_{c}^{D^+}(1-z) = d_c^d(z) = d_{\bar c} ^{\bar u} (z)   , \\
&  \hspace*{-3mm}\hat{d}_{u}^{c}(z) = \hat{d}_{u}^{\bar D^0}(1-z) =  \hat{d}_{u}^{D^- }(1-z) = d_d^c (z) = d_{\bar u} ^{\bar c} (z)    , \\
&  \hspace*{-3mm}\hat{d}_{c}^{s}(z) = \hat{d}_{c}^{D_s^+}(1-z)  , \\
&  \hspace*{-3mm}\hat{d}_{s}^{c}(z) = \hat{d}_{s}^{D_s^-}(1-z)  . 
\label{EQ:GparityF}
\end{align}
\end{subequations}
The full fragmentation functions~\eqref{EQ:REDF} also exhibit symmetry relations, 
\begin{subequations}
\begin{align}
\color{blue}{
D_{u}^{\pi^+}(z)}  \ &  \color{blue}{ = D_{\bar{d}}^{\pi^+}(z) = D_{\bar{u}}^{\pi^-}(z) = D_{d}^{\pi^-}(z)}  \,, \\
D_{u}^{\pi^-}(z) & = D_{\bar{d}}^{\pi^-}(z) = D_{\bar{u}}^{\pi^+}(z) = D_{d}^{\pi^+}(z) \,, \\
\color{blue}{D_u^{\pi^0}(z) } \ & \color{blue}{= D_d^{\pi^0}(z) = D_{\bar{u}}^{\pi^0}(z) = D_{\bar{d}}^{\pi^0}(z)} \,, \\
\color{blue}{ D_{u}^{K^+} \!(z)} \ &  \color{blue}{ = D_{d}^{K^0}(z) = D_{\bar{u}}^{K^-}(z) = D_{\bar{d}}^{\bar{K}^0}(z)} \,, \\
D_u^{K^0}(z) & =D_d^{K^{+}}(z)  =D_{\bar{u}}^{\bar{K}^0}(z) =D_{\bar{d}}^{K^{-}}(z) \,, \\
D_u^{K^{-}}(z) & = D_d^{\bar{K}^0}(z) = D_{\bar{u}}^{K^{+}}(z) = D_{\bar{d}}^{K^0}(z) \,, \\
D_u^{\bar{K}^0}(z) & = D_d^{K^{-}}(z) = D_{\bar{u}}^{K^0}(z) = D_{\bar{d}}^{K^{+}}(z) \,, \\
D_{u}^{D^0}(z) &= D_{d}^{D^+}(z) = D_{\bar{u}}^{\bar{D}^0}(z) = D_{\bar{d}}^{D^-}(z)\,,\\ 
\color{blue}{ D_{u}^{\bar{D}^0}(z) } \ & \color{blue}{ = D_{d}^{D^-}(z) = D_{\bar{u}}^{D^0}(z) = D_{\bar{d}}^{D^+}(z)} \,,\\ 
D_u^{D^+ }(z)  &=  D_d^{D^0}(z)  = D_{\bar u}^{D^-}(z) =  D_{\bar d}^{\bar D^0}(z)  \,,\\ 
D_u^{D^- }(z)  & = D_d^{\bar D^0} (z)  =  D_{\bar u}^{D^+}(z) =  D_{\bar d}^{D^0}(z)  \,, \\
D_{u}^{D_s^+}(z) &= D_{d}^{D_s^+}(z) = D_u^{D_s^-}(z) = D_d^{D_s^-}(z)\,,\\ 
\color{blue}{ D_{s}^{K^-}(z)} &= \color{blue}{ D_{s}^{\bar{K}^0}(z) = D_{\bar{s}}^{K^+}(z) = D_{\bar{s}}^{K^0}(z })\,, \\
D_{s}^{K^+}\! (z) &= D_{s}^{K^0}(z) = D_{\bar{s}}^{K^-}(z) = D_{\bar{s}}^{\bar{K}^0}(z)\,, \\
D_{s}^{D^-} \!(z) & = D_{s}^{D^+}(z)  = D_{s}^{D^0}(z)  = D_{s}^{\bar D^0}(z) \,,  \\
D_{s}^{\pi^+}(z) &= D_{s}^{\pi^-}(z) = D_{s}^{\pi^0}(z)\,, \\
D_{c}^{\pi^+}(z) &= D_{c}^{\pi^-}(z) = D_{c}^{\pi^0}(z)\,, \\
D_{c}^{K^+}(z) &= D_{c}^{K^0}(z)=D_{\bar{c}}^{K^-}(z)=D_{\bar{c}}^{\bar{K}^0}(z) \,,   \\ 
D_{c}^{K^-}(z) &= D_{c}^{\bar{K}^0}(z)=D_{\bar{c}}^{K^+}(z)=D_{\bar{c}}^{K^0} (z)\,,   \\ 
\color{blue}{ D_{c}^{D^0}(z)} &= \color{blue}{ D_{c}^{D^+}(z)\, = D_{\bar{c}}^{\bar{D}^0}(z) = D_{\bar{c}}^{D^-}(z)} \,, \\
D_{c}^{\bar D^0}(z) &=  D_{c}^{D^-}(z)\, = D_{\bar{c}}^{D^0}(z) = D_{\bar{c}}^{D^+}(z) \, , \\
\color{blue}{ D_{c}^{D_s^+} \!(z)} \  &\color{blue}{ = D_{\bar{c}}^{D_s^-}(z)} \,,   \\ 
\color{blue}{ D_{s}^{D_s^-} \!(z) } \ & \color{blue}{ =  D_{\bar{s}}^{D_s^+}(z) }\, ,  \\
D_{c}^{D_s^-}(z) & =  D_{\bar{c}}^{D_s^+} (z) \, , \\
D_{s}^{D_s^+}(z) & =  D_{\bar{s}}^{D_s^-} (z) \, ,
\label{EQ:recurrency}
\end{align}
\end{subequations}
where \emph{favored} fragmentation functions are highlighted in blue.  These relations reflect the underlying flavor SU(3) symmetry and charge 
conjugation invariance of the fragmentation process, while accounting for the distinct mass scales of charm-containing mesons. 

These twenty-five symmetry relations imply that one must solve a system of an equal number of coupled jet equations~\eqref{EQ:REDF}, which read,
\begin{subequations}
\begin{align}
D_{u}^{\pi^+}(z) & = \hat{d}_{u}^{\pi^+}(z) + \sum_{Q\in\{u,d,s,c\}} \int_z^1 \frac{dy}{y} \, \hat{d}_u^{Q}\left(\frac{z}{y}\right) D_Q^{\pi^+}(y)\,  ,
\label{EQ:DFqm1} 
\end{align}
\begin{align}
D_{u}^{\pi^0}(z) & = \hat{d}_{u}^{\pi^0}(z) + \sum_{Q\in\{u,d,s,c\}} \int_z^1 \frac{dy}{y} \, \hat{d}_u^{Q}\left(\frac{z}{y}\right) D_Q^{\pi^0}(y)\,  ,
\label{EQ:DFqm2}  \\
D_{u}^{K^+} (z) & =  \hat{d}_{u}^{K^+}(z) + \sum_{Q\in\{u,d,s,c\}} \int_z^1 \frac{dy}{y} \, \hat{d}_u^{Q}\left(\frac{z}{y}\right) D_Q^{K^+}(y)\,  ,
\label{EQ:DFqm3}  \\
D_{u}^{\bar{D}^0}(z) & = \hat{d}_{u}^{\bar{D}^0}(z) + \sum_{Q\in\{u,d,s,c\}} \int_z^1 \frac{dy}{y} \, \hat{d}_u^{Q}\left(\frac{z}{y}\right) D_Q^{\bar{D}^0}(y)\,  ,
\label{EQ:DFq4}  \\
D_{s}^{K^-} (z) & =  \hat{d}_{s}^{K^-}(z) + \sum_{Q\in\{u,d,c\}} \int_z^1 \frac{dy}{y} \, \hat{d}_s^{Q}\left(\frac{z}{y}\right) D_Q^{K^-}(y)\,  ,
\label{EQ:DFqm5}  \\
D_{s}^{D_{s}^-} (z)  & =  \hat{d}_{s}^{D_{s}^-}(z) + \sum_{Q\in\{u,d,c\}} \int_z^1 \frac{dy}{y} \, \hat{d}_s^{Q}\left(\frac{z}{y}\right) D_Q^{D_{s}^-}(y)\,  ,
\label{EQ:DFqm6}  \\
D_{c}^{D^0} (z) & =  \hat{d}_{c}^{D^0}(z) + \sum_{Q\in\{u,d,s,c \}} \int_z^1 \frac{dy}{y} \, \hat{d}_c^{Q}\left(\frac{z}{y}\right) D_Q^{D^0}(y)\,  ,
\label{EQ:DFqm7}   \\
D_{c}^{D_{s}^+} (z)  & =  \hat{d}_{c}^{D_{s}^+}(z) + \sum_{Q\in\{u,d,s,c\}} \int_z^1 \frac{dy}{y} \, \hat{d}_c^{Q}\left(\frac{z}{y}\right) D_Q^{D_{s}^+}(y)\, , 
\label{EQ:DFqm8}  
\end{align}
for the favored fragmentation functions and otherwise, 
\begin{align}
D_{u}^{\pi^-}(z) & =  \sum_{Q\in\{u,d,s,c\}} \int_z^1 \frac{dy}{y} \, \hat{d}_u^{Q}\left(\frac{z}{y}\right) D_Q^{\pi^-}(y)\,  ,
\label{EQ:DFqm9}  \\
D_{s}^{\pi^+}(z) & =  \sum_{Q\in\{u,d,c\}} \int_z^1 \frac{dy}{y} \, \hat{d}_s^{Q}\left(\frac{z}{y}\right) D_Q^{\pi^+}(y)\,  ,
\label{EQ:DFqm10}  \\
D_{c}^{\pi^+}(z) & =  \sum_{Q\in\{u,d,s,c \}} \int_z^1 \frac{dy}{y} \, \hat{d}_c^{Q}\left(\frac{z}{y}\right) D_Q^{\pi^+}(y)\,  ,
\label{EQ:DFqm11}  \\
D_{s}^{K^+}(z) & =  \sum_{Q\in\{u,d,c\}} \int_z^1 \frac{dy}{y} \, \hat{d}_s^{Q}\left(\frac{z}{y}\right) D_Q^{K^+}(y)\,  ,
\label{EQ:DFqm12}  \\
D_{c}^{K^+}(z) & =  \sum_{Q\in\{u,d,s,c\}} \int_z^1 \frac{dy}{y} \, \hat{d}_c^{Q}\left(\frac{z}{y}\right) D_Q^{K^+}(y)\,  ,
\label{EQ:DFqm13}   \\
D_{u}^{K^0}(z) & =  \sum_{Q\in\{u,d,s,c\}} \int_z^1 \frac{dy}{y} \, \hat{d}_u^{Q}\left(\frac{z}{y}\right) D_Q^{K^0}(y)\,  ,
\label{EQ:DFqm14}  \\
D_{u}^{\bar{K}^0}(z) & =  \sum_{Q\in\{u,d,s,c\}} \int_z^1 \frac{dy}{y} \, \hat{d}_u^{Q}\left(\frac{z}{y}\right) D_Q^{\bar{K}^0}(y)\,  ,
\label{EQ:DFqm15} \\
D_{u}^{K^-}(z) & =  \sum_{Q\in\{u,d,s,c\}} \int_z^1 \frac{dy}{y} \, \hat{d}_u^{Q}\left(\frac{z}{y}\right) D_Q^{K^-}(y)\,  ,
\label{EQ:DFqm16}   \\
D_{c}^{K^-}(z) & =  \sum_{Q\in\{u,d,s,c\}} \int_z^1 \frac{dy}{y} \, \hat{d}_c^{Q}\left(\frac{z}{y}\right) D_Q^{K^-}(y)\,  ,
\label{EQ:DFqm17}  
\end{align}
\begin{align}
D_{u}^{D^0}(z) & =  \sum_{Q\in\{u,d,s,c\}} \int_z^1 \frac{dy}{y} \, \hat{d}_u^{Q}\left(\frac{z}{y}\right) D_Q^{D^0}(y)\,  ,
\label{EQ:DFqm18}  \\
D_{u}^{D^+}(z) & =  \sum_{Q\in\{u,d,s,c\}} \int_z^1 \frac{dy}{y} \, \hat{d}_u^{Q}\left(\frac{z}{y}\right) D_Q^{D^+}(y)\,  ,
\label{EQ:DFqm19}  \\
D_{u}^{D^-}(z) & =  \sum_{Q\in\{u,d,s,c\}} \int_z^1 \frac{dy}{y} \, \hat{d}_u^{Q}\left(\frac{z}{y}\right) D_Q^{D^-}(y)\,  ,
\label{EQ:DFqm20}  \\
D_{s}^{D^0}(z) & =  \sum_{Q\in\{u,d,c\}} \int_z^1 \frac{dy}{y} \, \hat{d}_s^{Q}\left(\frac{z}{y}\right) D_Q^{D^0}(y)\,  ,
\label{EQ:DFqm21}  \\
D_{c}^{\bar{D}^0}(z) & =  \sum_{Q\in\{u,d,s,c\}} \int_z^1 \frac{dy}{y} \, \hat{d}_s^{Q}\left(\frac{z}{y}\right) D_Q^{\bar{D}^0}(y)\,  ,
\label{EQ:DFqm22}  \\
D_{u}^{D_{s}^-}(z) & =  \sum_{Q\in\{u,d,s,c\}} \int_z^1 \frac{dy}{y} \, \hat{d}_u^{Q}\left(\frac{z}{y}\right) D_Q^{D_{s}^-}(y)\, , 
\label{EQ:DFqm23}  \\
D_{s}^{D_{s}^+} (z)  & =  \sum_{Q\in\{u,d,c\}} \int_z^1 \frac{dy}{y} \, \hat{d}_s^{Q}\left(\frac{z}{y}\right) D_Q^{D_{s}^+}(y)\,  ,
\label{EQ:DFqm24}  \\
D_{c}^{D_{s}^-}\! (z) & =  \sum_{Q\in\{u,d,s,c\}} \int_z^1 \frac{dy}{y} \, \hat{d}_c^{Q}\left(\frac{z}{y}\right) D_Q^{D_{s}^-}(y)\,  .
\label{EQ:DFqm25}  
\end{align}
\end{subequations}
Note that the splitting functions $\hat d_s^s (z/y)$ are not summed over, as we do not consider $\bar ss$ states via $\eta-\eta'$ mixing, for instance. 
On the other hand, the $\hat d_c^c (z/y) = \hat d_c^{\eta_c} (1-z/y)$ splitting function can be taken into account, though its effect is negligible
as can be inferred from Fig.~\ref{fig:splitting_etac}. Adding a term $\hat d_c^{\eta_c}(z)$ to Eq.~\eqref{EQ:NEFF} has hardly any effect on the normalization
of $\hat{d}_{c}^{D^{0}}(z)$  and $\hat{d}_{c}^{D_{s}^{+}}(z)$. 


\begin{table}[b!]
\centering 
\renewcommand{\arraystretch}{1.5}
\setlength{\tabcolsep}{6pt}
\begin{tabular}{c|c|c|c}
\hline\hline
     $\hat{d}(z)^m _q$ & $N$ & $\alpha$ & $\beta$ \\ \hline
    $\hat{d}(z)^{\pi^{0}}_{u}$  & 5.91 & 2,41 & 1.38  \\ 
    $\hat{d}(z)^{\pi^{+}}_{u}$           & 11.83 & 2.41 & 1.38 \\
   $\hat{d}(z)^{K^{+}}_{u}$            & 53.60 & 5.35 & 2.40  \\ 
  $\hat{d}(z)^{\bar{D}^{0}}_{u}$            & $67.23 \times 10^{3}$ & 12.92 & 7.04  \\ 
  $\hat{d}(z)^{K^{-}}_{s}$             & 51.73 & 4.74 & 1.75 \\
$\hat{d}(z)^{D^{-}_{s}}_{s}$         & $26.48\times 10^{4}$& 13.89 & 7.53  \\
$\hat{d}(z)^{D^{0}}_{c}$         & $58.12 \times 10$ & 10.91 & 2.24 \\
$\hat{d}(z)^{D^{+}_{s}}_{c}$    & $22.10\times 10^{4}$ & 18.26 & 4.96  \\
\hline\hline
\end{tabular}
\caption{\label{tab:Nparameters} Parameters of the normalized fragmentation function in Eq.~\eqref{EQ:Neff_fit}.}   
\vspace*{-4mm}
\end{table} 


\begin{figure}[t!] 
\vspace*{-7mm}
\hspace*{-7mm}
\centering
  \includegraphics[scale=0.85,angle=0]{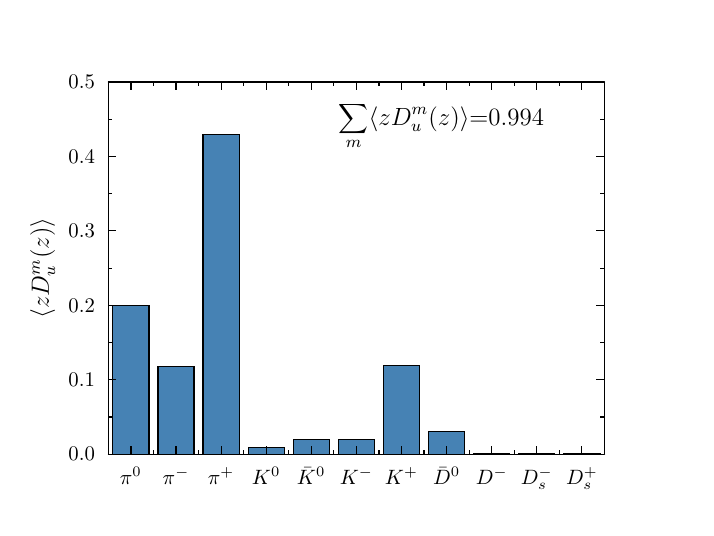} 
  \vspace*{-1.1cm}
\caption{Distribution of momentum fractions carried by a meson $m$ produced by an initial up quark. Note that the moments for  fragmentation 
into $D^+$ and $D^0$ are consistent with zero and therefore ignored in the chart.  }
\label{fig:Dumz} 
\end{figure}

\begin{figure}[h!] 
\vspace*{-7mm}
\hspace*{-7mm}
\centering
  \includegraphics[scale=0.85,angle=0]{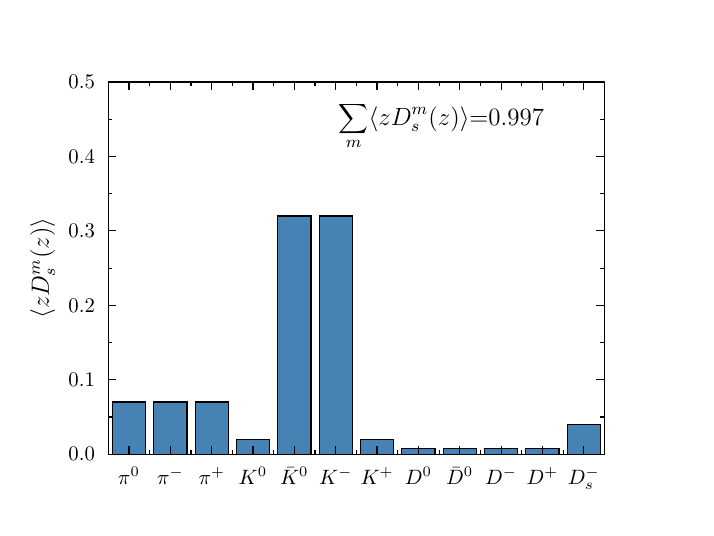} 
  \vspace*{-1.1cm}
\caption{Momentum fraction distributions carried by a meson $m$ produced by an initial strange quark. The moment of $D_s^{D^+}$ is much smaller than
that of  $s\to D^+, D^-, D^0, \bar D^0$ fragmentations.}
\label{fig:Dsmz} 
\end{figure}


\begin{figure}[t!] 
\vspace*{-7mm}
\hspace*{-7mm}
\centering
  \includegraphics[scale=0.85,angle=0]{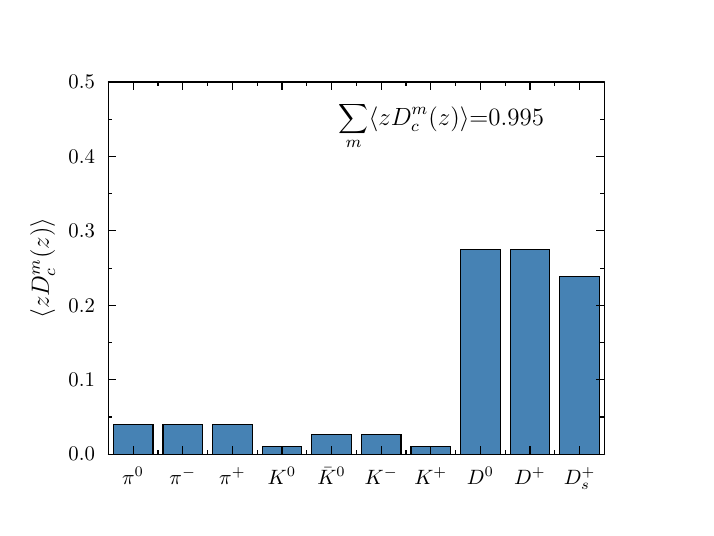} 
  \vspace*{-1.1cm}
\caption{Momentum fraction distributions carried by a meson $m$ produced by an initial charm quark. The moments of the $c\to D^-$ and $c\to D_s^-$ fragmentations
are negligibly small. }
\label{fig:Dcmz} 
\end{figure}

\section{Results}

With the numerical solutions of the fragmentation functions, Eqs.~\eqref{EQ:DFqm1} to \eqref{EQ:DFqm25}, obtained by an iterative process,
we compute the momentum fractions $\langle z D_q^m(z) \rangle$ carried by mesons of type $m$ in jets originating from $u$, $s$, and $c$ quarks.
This extends an earlier study in the Nambu--Jona-Lasinio model~\cite{Matevosyan:2010hh} to include charm quarks. The moments of the 
fragmentation functions are displayed as column charts in Figs.~\ref{fig:Dumz}, \ref{fig:Dsmz} and \ref{fig:Dcmz} and corroborate the momentum 
sum rule,
\begin{equation}
  \sum_m \langle z D_q^m(z) \rangle = 1 \, , 
   \label{EQ:sumDzm}
\end{equation}
within the numerical precision of our calculations, which is less than $1\%$. The results demonstrate consistent conservation across all quark flavors, 
including the additional charm sector.

Software packages that implement numerically the Dokshitzer-Gribov-Lipatov-Altarelli-Parisi  (DGLAP) evolution equations often rely on an analytic
representation of the parton distribution functions or the elementary fragmentation functions. The numerical fragmentation functions introduced in Eqs.~\eqref{EQ:GparityI} 
to \eqref{EQ:GparityF} can be reproduced with the algebraic form,
\begin{align}
  \hat{d}(z)_{q}^{m} = N z^{\alpha}(1-z)^{\beta},
  \label{EQ:Neff_fit}
\end{align}
with $N$, $\alpha$, and $\beta$ denoting fit parameters. The parameter sets of Eq.~\eqref{EQ:Neff_fit} are listed in the table~\ref{tab:Nparameters}.

\begin{figure}[t!] 
\vspace*{-5mm} \hspace*{-4mm}
  \includegraphics[scale=0.54,angle=0]{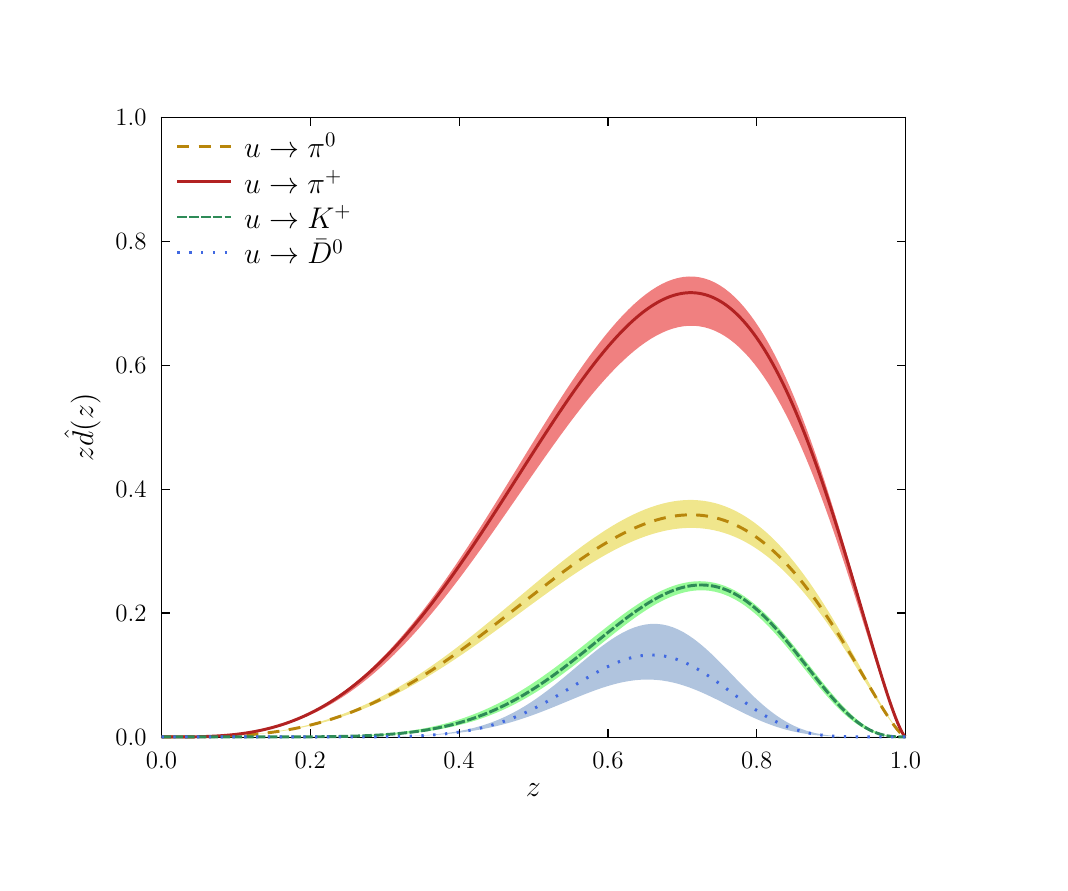} 
  \vspace*{-1.2cm}
\caption{The normalized fragmentation functions $\hat{d}_{u}^{m}(z)$ of the elementary  fragmentations  $u \to \pi^{0}$, $\pi^{+}$, $K^{+}$, $\bar{D}^{0}$. }
\label{fig:splitting_u} 
\end{figure}

\begin{figure}[t!] 
\vspace*{-5mm} \hspace*{-4mm}
  \includegraphics[scale=0.54,angle=0]{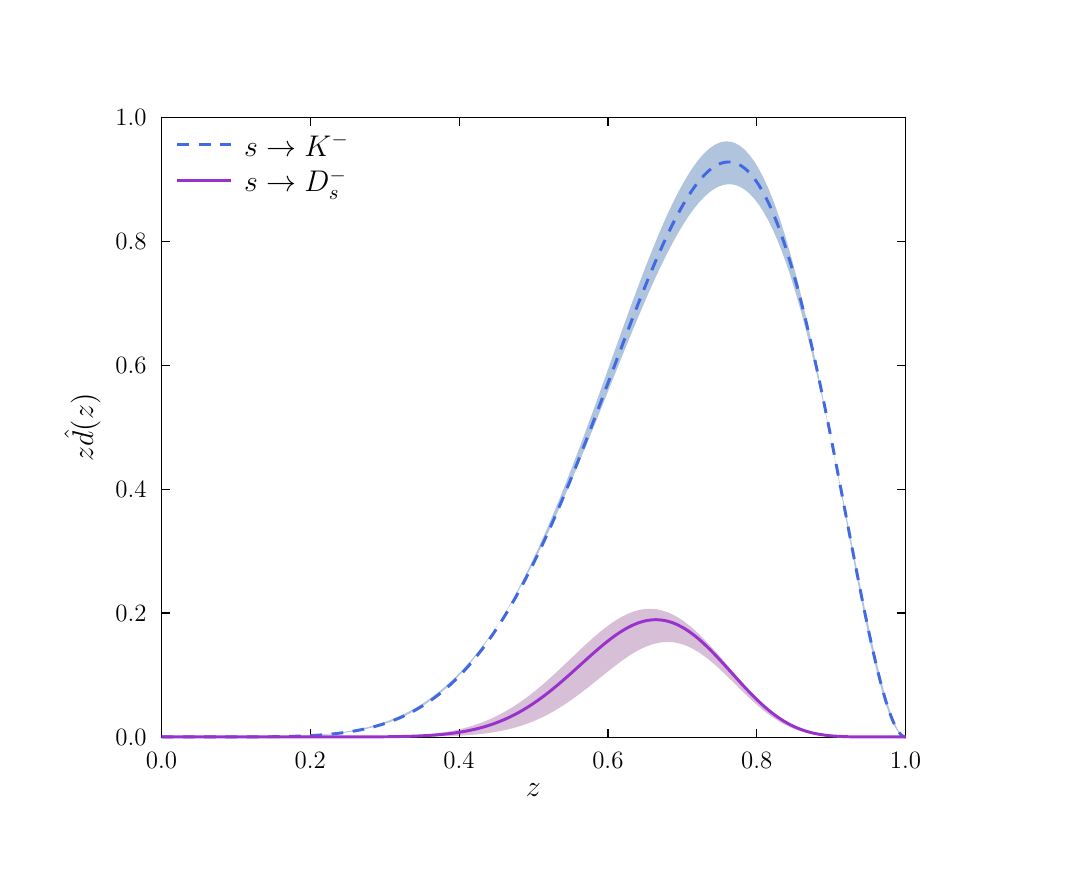} 
  \vspace*{-1.2cm}
\caption{The normalized fragmentation functions $\hat{d}_{s}^{m}(z)$ of the elementary  fragmentations $s\to K^{-}$, $D^{-}_{s}$}
\label{fig:splitting_s} 
  \vspace*{-3mm}
\end{figure}

\begin{figure}[t!] 
\vspace*{-3mm} \hspace*{-4mm}
  \includegraphics[scale=0.54,angle=0]{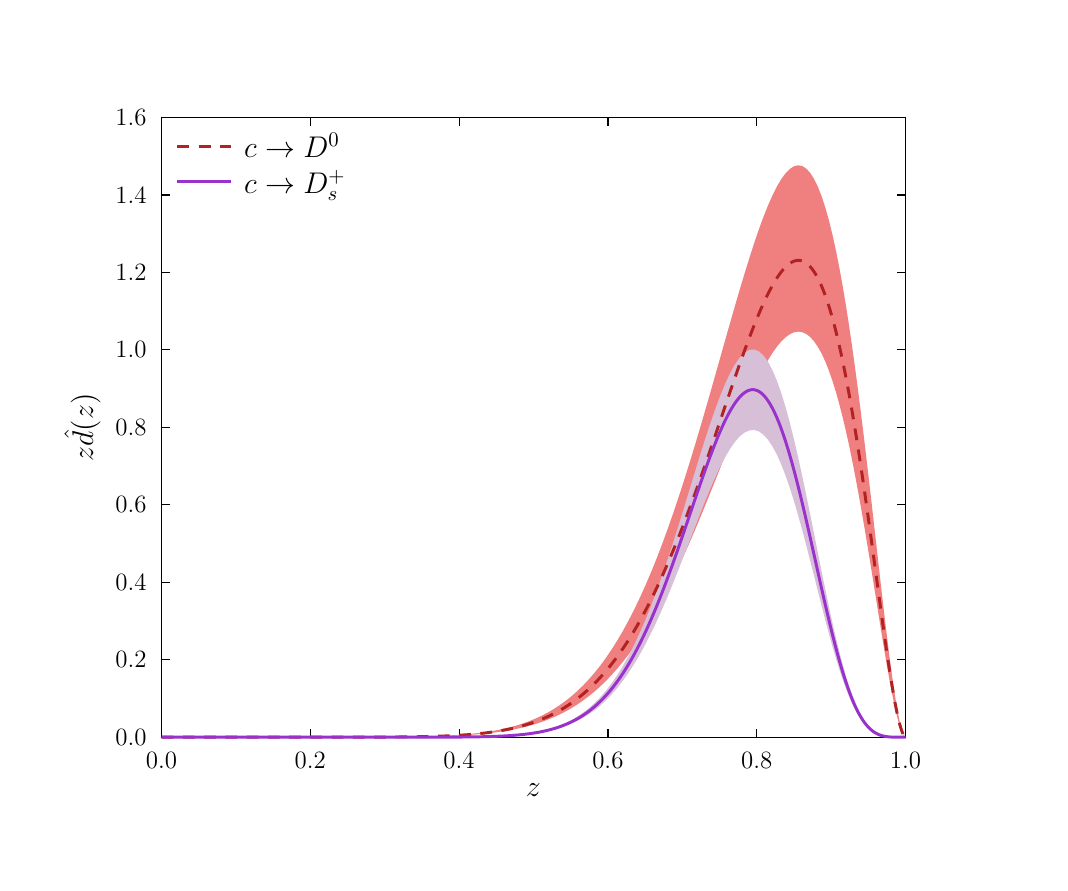} 
  \vspace*{-1.2cm}
\caption{The normalized  fragmentation  functions $\hat{d}_{c}^{m}(z)$ of the elementary  fragmentations $c\to D^{0}$, $D^{+}_{s}$}
\label{fig:splitting_c} 
 \vspace*{-2mm}
\end{figure}

\begin{figure}[t!] 
\vspace*{-5mm} \hspace*{-4mm}
  \includegraphics[scale=0.54,angle=0]{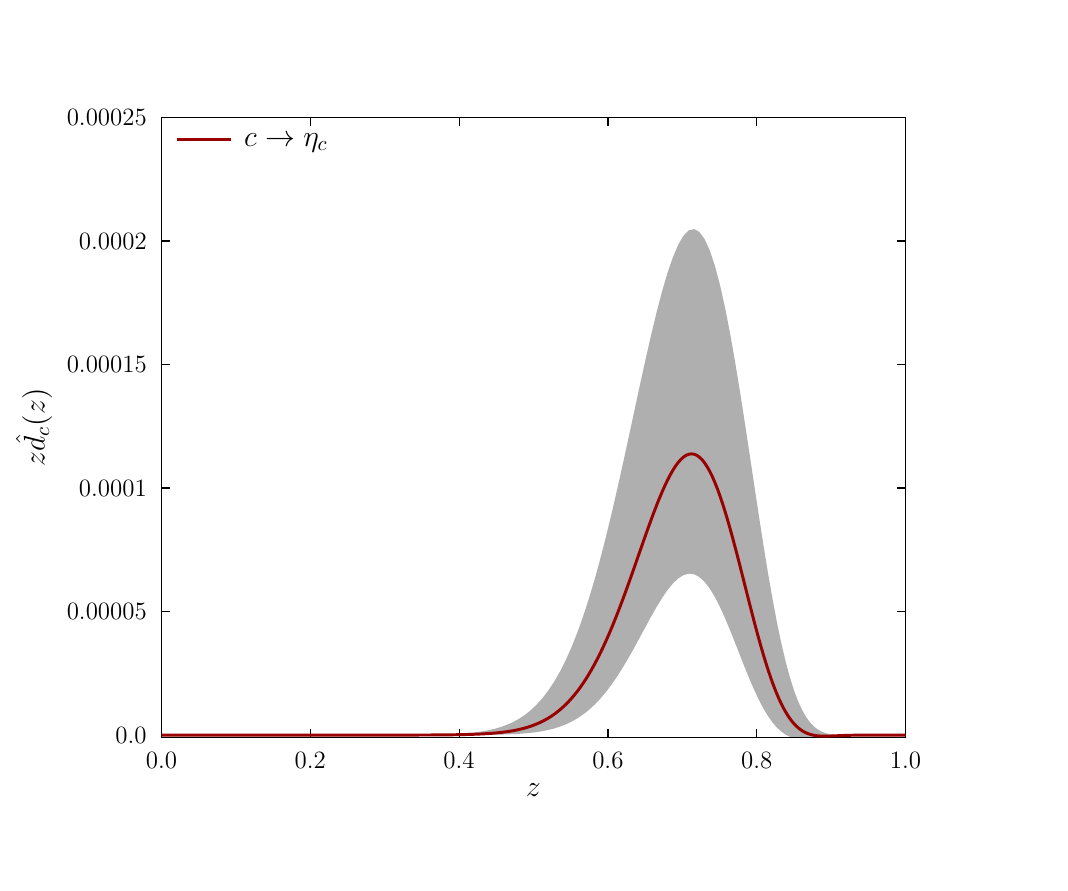} 
  \vspace*{-1cm}
\caption{The normalized  fragmentation function $\hat{d}_{c}^{\eta_c}(z)$ is strongly suppressed with respect to $\hat d_s^{D_s^-}(z)$ and $\hat d_c^{D_s^+}(z)$, 
Figs.~\ref{fig:splitting_s} and \ref{fig:splitting_c}, respectively. }
\label{fig:splitting_etac} 
\end{figure}

The elementary fragmentation functions displayed in Figs.~\ref{fig:splitting_u} and \ref{fig:splitting_s} clearly exhibit the expected hierarchy of 
hadron emission probabilities. Namely, the fragmentation of light and strange quarks into $D$ mesons is significantly suppressed with respect the 
fragmentation into pions and kaons  throughout the entire $z$-range. These results are also consistent with the full pion fragmentation function
of an earlier study~\cite{daSilveira:2024ddq}. The shaded error bands throughout all figures arise from the uncertainties in fitting numerical solutions 
of the BSA to spectral Nakanishi-like representations which allow for analytic calculations, as discussed in detail in Appendix A of Ref.~\cite{daSilveira:2024ddq}.

On the other hand, Fig.~\ref{fig:splitting_c} shows that the  $c \rightarrow D^0$ and $c \rightarrow D_s^+$ fragmentation functions are qualitatively compatible, 
though the latter is somewhat suppressed and peaks at a lower momentum fraction. This behavior reflects the small mass difference between the two charmed mesons,
\emph{i.e.\/} the charm is more likely to fragment into a $D^0$ meson. Moreover, the magnitudes of the $s \to D_s^-$ and $c \to D_s^+$ fragmentations, 
Figs.~\ref{fig:splitting_s} and ~\ref{fig:splitting_c} respectively,  differ considerably and reflect the on-shell condition of the cut diagram in Eq.~\eqref{EQ:fragLF}: 
in the former case the charm quark is on shell, whereas in the latter case the strange quark is on shell. When both quarks are a charm, this on-shell effect combined 
with the narrowness of the charmonium's BSA exacerbates the suppression, as becomes clear from Fig.~\ref{fig:splitting_etac}.

\begin{figure}[t!] 
\vspace*{-8mm} \hspace*{-4mm}
  \includegraphics[scale=0.54,angle=0]{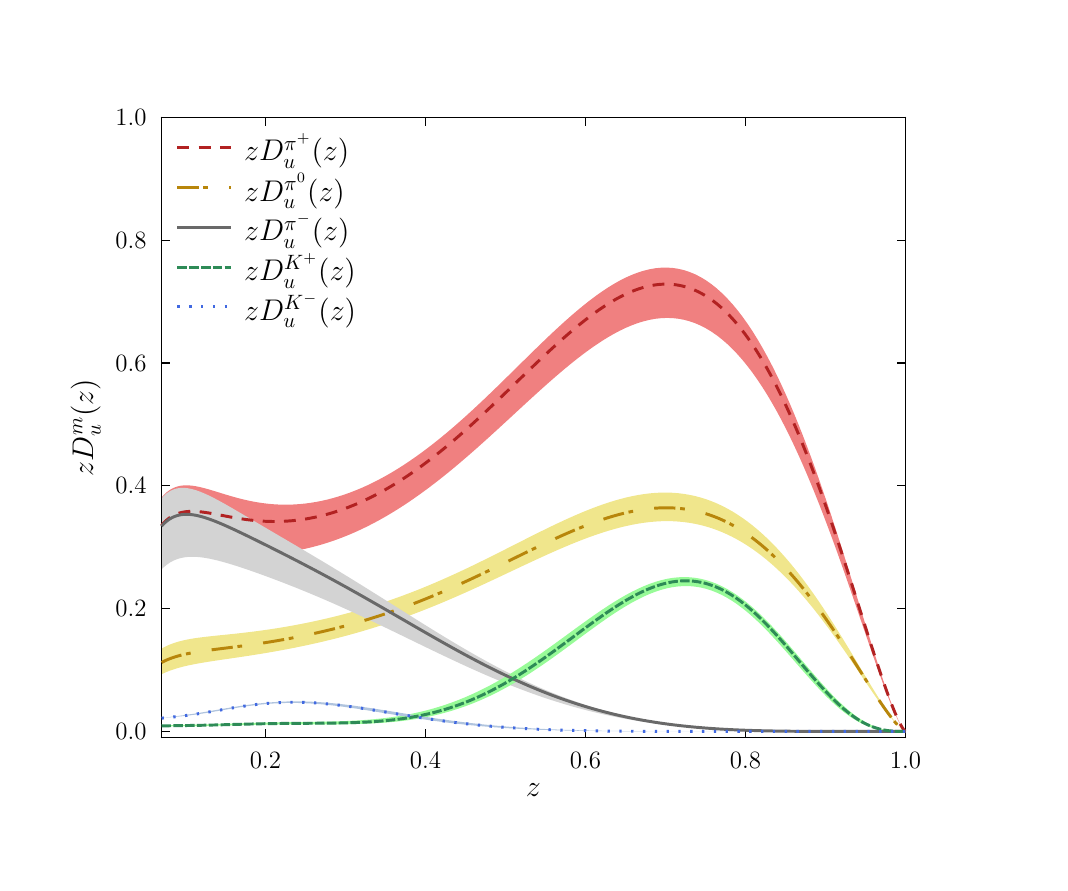} 
  \vspace*{-1.2cm}
\caption{Full fragmentation functions $zD_u^m(z)$ of an up quark into pions and kaons}
\label{fig:zDf_u}
\end{figure}

\begin{figure}[t!] 
\vspace*{-2mm} \hspace*{-4mm}
  \includegraphics[scale=0.54,angle=0]{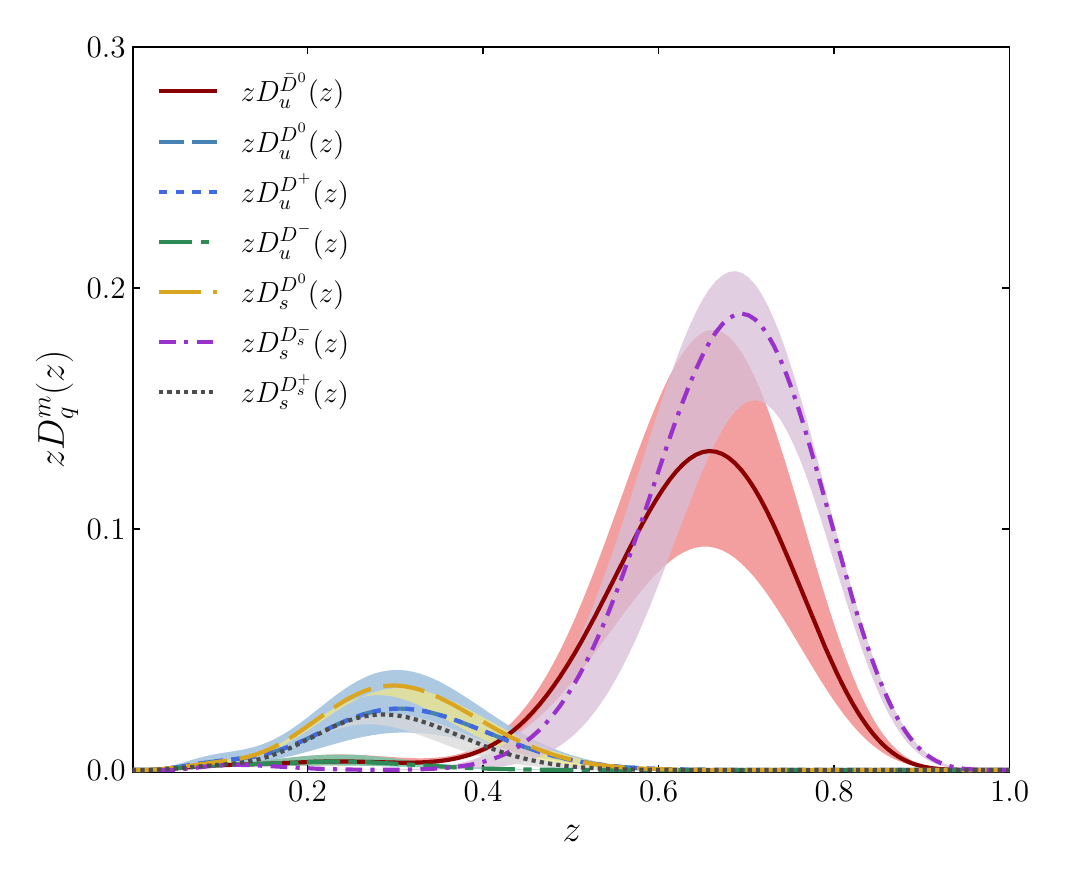} 
  \vspace*{-1.2cm}
\caption{Full fragmentation functions $zD_u^m(z)$ and  $zD_s^m (z)$ of up  and strange quarks into charmed mesons. }
\label{fig:zDf_usDmeson} 
\end{figure}

The full fragmentation functions $zD_u^m (z)$ are presented in Figs.~\ref{fig:zDf_u} and \ref{fig:zDf_usDmeson}. The dominant channels correspond to 
pion emission, with  $u \rightarrow \pi^+$ and $u \rightarrow \pi^0$ exhibiting the largest values across most of the $z$-range. Fragmentation into kaons 
is visibly suppressed for $z\lesssim 0.5$, and so is the emission into $D$ mesons, even in the case of the favored fragmentation channel $u\to \bar D^0$, 
as illustrated in  Fig.~\ref{fig:zDf_usDmeson}.  These findings are consistent with the expected mass hierarchy in the fragmentation process.

\begin{figure}[t!] 
\vspace*{-5mm} \hspace*{-4mm}
  \includegraphics[scale=0.54,angle=0]{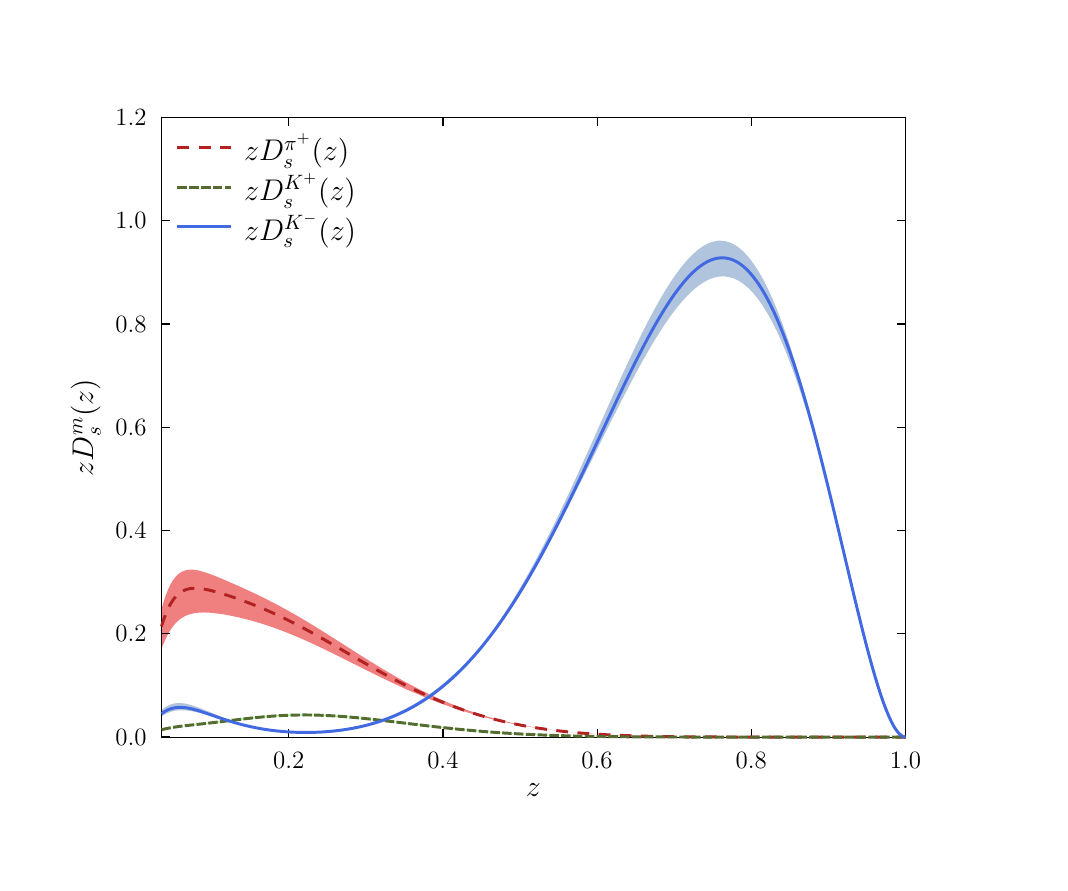} 
  \vspace*{-1.2cm}
\caption{Full fragmentation functions $zD_s^m(z)$ of a strange quark into pions and kaons.}
\label{fig:zDf_s} 
\end{figure}

\begin{figure}[t!] 
\vspace*{-5mm} \hspace*{-4mm}
\centering
  \includegraphics[scale=0.54,angle=0]{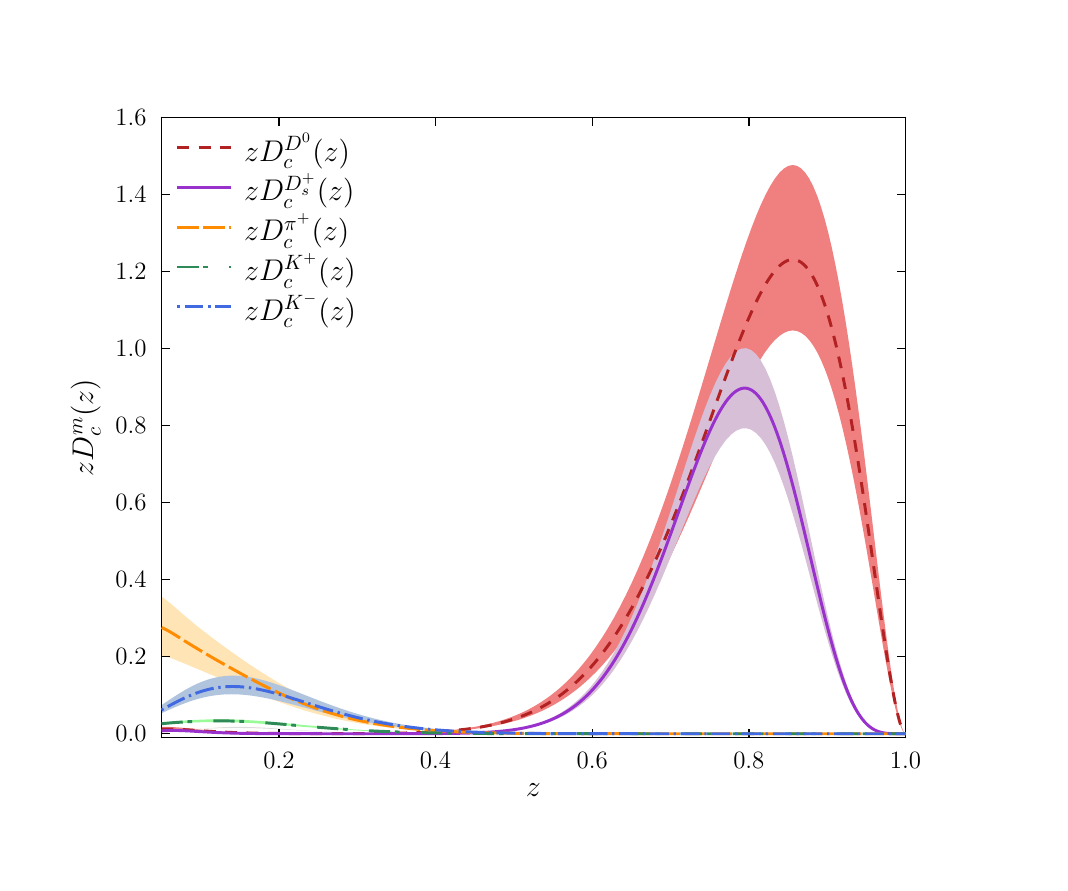} 
  \vspace*{-1.2cm}
\caption{Full fragmentation functions $zD_c^m(z)$ of a charm quark into pions, kaons and $D$ mesons.}
\label{fig:zDf_c} 
\end{figure}

The fragmentation functions of a strange quark in Fig.~\ref{fig:zDf_s} are dominated by the favored $s \rightarrow K^-$ emission that peaks at about $z\approx 0.75$, 
while the  $s \rightarrow K^+$  and $s \rightarrow \pi^\pm, \pi^0$  fragmentations are non-existing for $z\gtrsim 0.6$ and have modest support at low $z$. Likewise, 
as can be read from Fig.~\ref{fig:zDf_usDmeson}, the $s \rightarrow D^0$ and $s \rightarrow D_s^+$ fragmentation functions exhibit a broad yet much suppressed peak at  
$z\approx 0.3$ and vanish otherwise. On the other hand, the channels $s \rightarrow \bar D^0$ and $s \rightarrow D_s^-$ are favored at the elementary level and therefore 
also the corresponding full fragmentation functions. 

The charm fragmentation functions in Fig.~\ref{fig:zDf_c} are almost completely dominated by the favored channels, namely $c\rightarrow D^0$ (equal to
$c\rightarrow D^+$) and $c\rightarrow D^{D_s^+}$, with prominent narrow peaks centered at  intermediate-to-high  momentum fractions $z$, whereas the fragmentation 
into pions and kaons is strongly suppressed. Both functions are comparable in magnitude and shape, which is consistent with their small mass difference.

Fragmentation functions are scale dependent objects and the scale of our calculation was determined in Ref.~\cite{daSilveira:2024ddq} by matching the first moment 
of the valence quark distribution extracted from a $\pi N$ Drell-Yan analysis, $2 \langle x \rangle_v = 0.47(2)$~\cite{Sutton:1991ay,Gluck:1999xe} using next-to-leading order 
DGLAP evolution~\cite{Botje:2010ay} from $Q = 2\,$GeV to our model scale including $N_f=4$ thresholds. The latter was found to be $Q_0 = 0.63\;$GeV. 

\begin{figure}[t!] 
\vspace*{-5mm}
\hspace*{-4mm}
  \includegraphics[scale=0.54,angle=0]{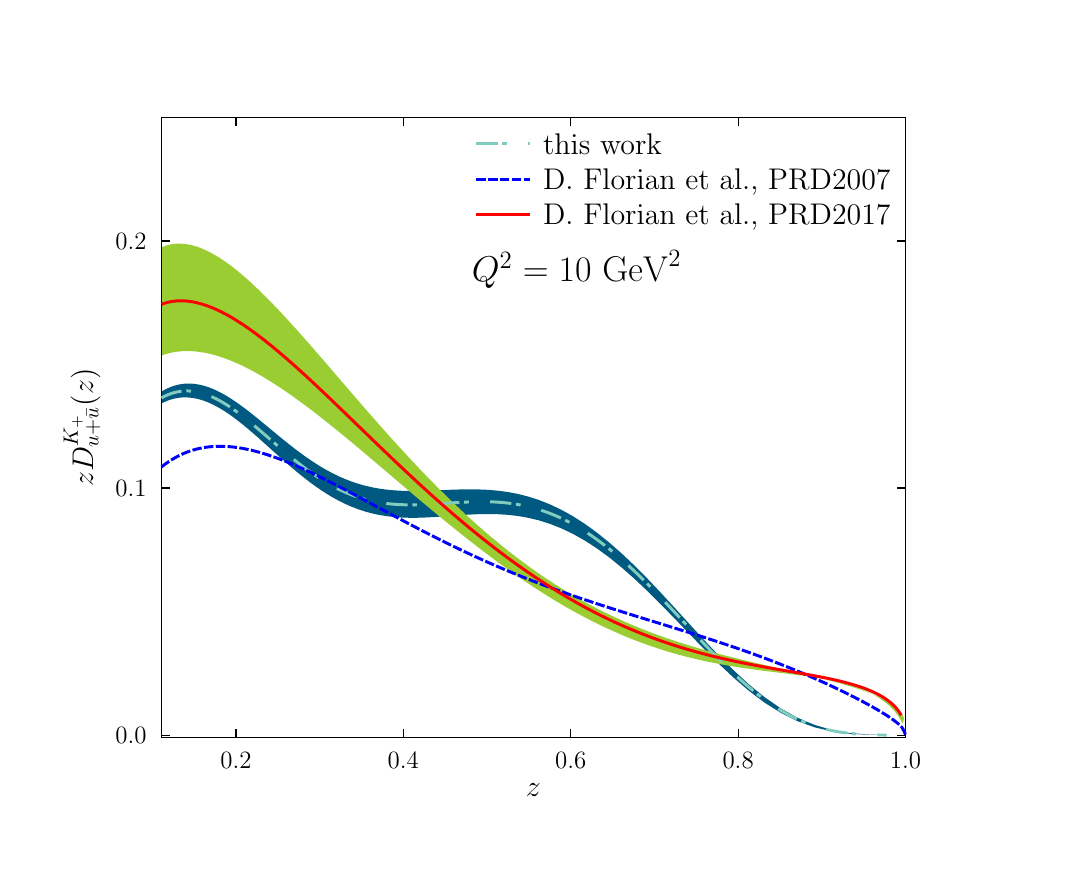} 
  \vspace*{-1.2cm}
\caption{The $zD_{u+\bar u}^{K^+} (z)$ fragmentation function for positively charged kaons evolved to the  scale $\mu = 10$\;GeV$^2$ (dashed-dotted curve). 
The green-shaded band depicts the global analysis extraction of Ref.~\cite{deFlorian:2017lwf} and the dashed blue curve describes the result of an earlier next-to-leading order
global analysis~\cite{deFlorian:2007aj}. }
\label{fig:zDuubarK} 
\end{figure}

Knowing the model scale allows us to use the related evolution kernels of the fragmentation functions~\cite{Hirai:2011si} including $N_f=4$ thresholds.
The $u+\bar u \to K^+$  fragmentation function evolved to 10\;GeV$^2$ is plotted in Fig.~\ref{fig:zDuubarK} along with the results of two global 
analyses~\cite{deFlorian:2007aj,deFlorian:2017lwf}. Our prediction reveals a rapidly increasing fragmentation function down to $z\approx 0.6$,
followed by a plateau and a further rise below $z\approx 0.3$, and compares reasonably well with the extracted fragmentation functions of Ref.~\cite{deFlorian:2017lwf} 
and \cite{deFlorian:2007aj} for momentum fractions $0\leq z \lesssim 0.5$. The extracted fragmentation functions, on the other hand, are monotonically rising functions 
which only saturate at low $z$-values.


\section{Conclusions}

In this work, we presented a unified, Poincar\'e-covariant calculation of fragmentation functions for light and heavy–light pseudoscalar mesons, including the $\pi$, $K$, $D$, 
and $D_s$ mesons, within the nonperturbative framework of functional continuum Schwinger equations. Starting from fully dressed quark propagators and BSAs obtained in 
rainbow-ladder truncation, we computed elementary quark-to-meson fragmentation functions and embedded them in a set of twenty-five coupled integral  jet equations 
that consistently resum the cascade of all possible hadronizations into the mesons considered in this work.

The resulting fragmentation functions satisfy the appropriate normalization and momentum sum rules with high numerical accuracy and exhibit the expected hierarchy 
dictated by quark masses and flavor structure. Fragmentation of light quarks into charmed mesons is strongly suppressed, while charm fragmentation is dominated by 
the favored $c \to D$ and $c \to D_s$ channels at intermediate and large momentum fractions. The relative magnitudes of the heavy-light fragmentation 
functions follow from the underlying quark mass dependence, the on-shell constraints of the cut diagram, and the momentum distributions encoded in the mesonic BSAs.

After determining the model scale independently, we evolve the $u\to K^+$ and $\bar u\to K^+$ fragmentation functions using DGLAP evolution and find reasonable agreement 
with existing global analyses,  in particular for kaon production at moderate values of $z$. This demonstrates that a covariant bound-state description of mesons, combined 
with a jet-cascade formalism,  can provide phenomenologically relevant fragmentation functions without introducing ad-hoc parametrizations or point-like approximations 
for the meson's wave functions.

Overall, our results establish a consistent and predictive framework for studying quark hadronization across light and heavy sectors within a unified  
framework. Our approach can be systematically extended to include vector mesons, baryons, and polarized fragmentation functions, and offers a natural bridge 
between continuum QCD methods and phenomenological applications in high--energy scattering processes.


\section{Acknowledgments}

We are grateful to Shunzo Kumano for providing the evolution code of the fragmentation functions and very helpful explanations. 
This work received financial support from the S\~ao Paulo Research  Foundation (FAPESP), grant no.~2023/00195-8 and from the National Council for Scientific 
and Technological Development (CNPq), grant no.~409032/2023-9. R.\,C.\,S  is supported by a CAPES fellowship, grant no.~88887.086709/2024-00.



\end{document}